\documentstyle[12pt]{article}
\def\lo{\langle 0 |}
\def\ro{ | 0 \rangle }

\def\gmf{\gamma _{5}}

\def\la{\langle }
\def\ra{ \rangle }

\newcommand{\beq}{\begin{equation}}
\newcommand{\eeq}{\end{equation}}
\newcommand{\bea}{\begin{eqnarray}}
\newcommand{\eea}{\end{eqnarray}}

\begin{document}
 \begin{titlepage}
\begin{flushright}
hep-ph/9807335
\end{flushright}
\vskip1.8cm
\begin{center}
 {\LARGE  Axion Potential, Topological Defects
\vskip0.5cm 
and CP-odd Bubbles in QCD} 
\vskip1.0cm
{\Large Igor~Halperin} 
and 
{\Large Ariel~Zhitnitsky}
\vskip0.5cm
        Physics and Astronomy Department \\
        University of British Columbia \\
        6224 Agricultural Road, Vancouver, BC V6T 1Z1, Canada \\ 
        {\small e-mail: 
higor@physics.ubc.ca \\
arz@physics.ubc.ca }\\
\vskip0.5cm
{\it Extended version of the talk given by AZ at the Axion workshop
at Gainesville, FL, March 13-15,1998 }
\vskip1.5cm
{\Large Abstract:\\}
\end{center}
\parbox[t]{\textwidth}{ 
It follows on general grounds that
the $\theta$ dependence in QCD is more complicated than 
suggested by  the large $N_c$ approach or instanton
arguments. 
Generically, the vacuum energy $E_{vac}(\theta)$ 
is a  multi-valued function of $ \theta $ admitting  
the existence of metastable states. We discuss decays
of such metastable vacua in the theory with and without
the axion, and point out the potential
relevance of this and related phenomena
for constraining a dark matter axion. Based on the 
analysis of the axion potential, an idea for a  
new axion search experiment at RHIC is suggested. It is 
noted that the false vacuum decay proceeds with 
maximal violation of CP, even if $ \theta = 0 $. We 
further speculate that the famous Sakharov criteria for 
baryogenesis could be satisfied at the  
%In the context of the 
%axion physics, this may suggest new mechanisms of the 
%axion production. Potentially, these phenomena  
%may lead to the necessity  to reconsider the  constraints on a dark
%matter axion. Based on the analysis of the axion potential $V(a)$, 
%we suggest a new idea for the axion search experiment at RHIC. As a 
%byproduct of our analysis,
%we point out that two out of three Sakharov's criteria
%for the baryogenesys are satisfied in our scheme, and 
%speculate on the possibility for the baryogenesis at the
QCD scale.  }
\vspace{1.0cm}
\end{titlepage}

 \section{Introduction}

The axion is perhaps the most well-motivated  dark matter 
candidate ever
since the early 1980's. Originally, 
the axion \cite{axion}
and its invisible modifications \cite{DFSZ}
 has been introduced into the theory
 not as a dark matter candidate, but rather as the particle 
which  solves an internal  fundamental QCD problem --the strong 
CP problem. Twenty years later we
 still do not know a better solution of this problem,
therefore, we should admit that the axion solution of 
the strong CP problem has successfully passed
the timing test \cite{review}. In this respect, unlike many other  
exotic particles, the  axion is   the unique 
dark matter candidate.

In the standard thermal scenario, cosmic axions can be created
through the radiative decay of the axion strings \cite{strings} or 
due to the
`misalignment' effect \cite{misal} at the QCD phase transition, 
see e.g. \cite{review}. These scenarios do not exhaust
all possible mechanisms for the axion production. In particular, 
 one more 
possible scenario for the axion production will be suggested
below in the text. 
We note that constraints  on the axion mass (or the coupling constant
$f_a\sim m_a^{-1}$) are very sensitive to the specific scenario 
for production 
of the cosmic axions, 
inflationary model, ratio $ f_{a}/T_{reh} $ (where
$ T_{reh} $ is the reheating temperature   of
 the universe at
 the end of inflation), etc. We restrict ourselves by considering
possible  constraints which follow from the QCD part of the problem.

We start with recalling some popular ans\"{a}tze for the 
axion potential. In 
the  `misalignment' mechanism, after the axion mass switches on
at the QCD scale, the 
axion field
begins to coherently oscillate about the minimum at $ \theta = 
0 $. Such a picture corresponds to the simplest possible 
choice for
the axion potential $V(a)= m_a^2 a^2/2 $ with
possible anharmonic corrections, and does
 not take into account the periodic properties of the axion field
$a\rightarrow a+2\pi f_a $. One can do better and 
consider potential $V(a)\sim \cos(a/f_a)$ motivated by 
instantons,
which is a periodic function of $a$.  One can calculate
 anharmonic corrections for fields near the top of the 
potential \cite{turner},
and account for topological effects
due to the  appearance of domain walls \cite{strings}
for this potential. All these  
effects are potentially
extremely important,  and  can drastically change
the whole picture of the cosmic axion production. 
 
However,  
the above potentials $V(a)$ are the model expressions
not derived from QCD. 
Moreover, they do not satisfy some general
requirements of the theory, see Sect.2,3. Therefore, the corresponding
calculations should be considered as qualitative estimates
of possible phenomena which may happen in the development of 
the axion field.
 
We recall at this point  that
there is the one-to-one correspondence between the 
form of the axion potential
$V(a)$  and the 
vacuum energy $E_{vac}(\theta)$ 
as a function of the  fundamental QCD parameter $\theta$.
Indeed, the axion solution of the strong CP problem suggests 
that $\theta$ parameter in QCD is promoted
to the  dynamical axion field $\theta \rightarrow a(x)/f_a $, and 
the QCD vacuum energy $E_{vac}(\theta)$ 
 becomes the axion potential $V(a)$. Therefore, 
our problem of analysing $V(a)$ amounts to the 
study of $E_{vac}(\theta)$ in QCD without the axion.

Recently, there has been a progress in understanding the 
general properties of the $ \theta $ dependence in QCD \cite{HZ}
based on a new development \cite{KS} in supersymmetric (SUSY) theories
(Sect.4 and 5).
The purpose of this paper is to discuss the implications
of this new understanding for the study of local
and global properties of the axion potential. 
As for the former (Sect.6.1), we 
note that the temperature dependence of 
the axion mass (and of entire axion potential) 
can be related with that of
the QCD vacuum quark and gluon
condensates whose temperature dependence  
is understood (from lattice or model calculations).
Yet, the most interesting phenomena are related to the 
global properties of the axion potential (Sect.6.2).  
As will be discussed
below,
the vacuum energy $E_{vac}(\theta)$ (and consequently, the axion 
potential $V(a)$) 
is generically multi-valued, i.e.
in some parametrical regions of  $\theta$         
extra local minima with higher energy could exist. It may lead 
to the phenomenon of the false vacuum decay through bubble 
nucleation. 
This process may provide a new mechanism for the axion production,
as, when the axion is present, it can be supplemented by rolling
to  $ \theta = 0 $ with emission of axions. 
 Potentially,
an account  of these and other related 
phenomena may lead to the 
necessity  to reconsider the  constraints on a dark
matter axion. 
Based on the analysis of the axion potential, we further suggest 
an idea for the new axion search experiment at RHIC
(Sect.7). 
Last but not least, we note that because CP is 100 \% violated in such
false vacuum decays, the famous Sakharov
criteria \cite{Sakh}
for baryogenesis could be satisfied
at the QCD scale (Sect.8).

 \section{What is known about $E_{vac}(\theta)$?  }

As we already mentioned, our problem is reduced to the analysis
of the $\theta$ dependence of the vacuum energy due to the exact 
correspondence
$\theta\rightarrow a/f_a,~~
E_{vac}(\theta)\rightarrow V(a/f_a)$. What do we know about
$E_{vac}(\theta)$? We know a few exact statements:\\
1. We know the exact Ward Identity (WI) which 
(near the chiral limit $m_q\rightarrow 0$
with equal masses ) takes the form \cite{WI}:
\beq
\label{1}
i \int dx \lo T \left\{ \frac{\alpha_s}{8 \pi} 
G \tilde{G} (x) ~
\frac{\alpha_s}{8 \pi} G \tilde{G} (0) \right\}\ro 
= - \frac{ \partial^2 E_{vac}(\theta)}{\partial \theta^2} =
\frac{m_q}{N_f}\lo\bar{\Psi}\Psi\ro +O(m_q^2).
\eeq
Therefore, at small $\theta$ the vacuum energy   is fixed
$E(\theta)-E(0)\sim f_{\pi}^2m_{\pi}^2\theta^2$ and, thus the
axion mass $m_a^2$ is also fixed: $V(a)-V(0)\sim f_{\pi}^2m_{\pi}^2
(a/f_a)^2 ,
~m_a^2\sim f_{\pi}^2m_{\pi}^2/ f_a^2$.
As it should be, in the chiral limit   the vacuum energy
does not depend on $\theta$.\\
2. The vacuum energy $E_{vac}(\theta)$ is a periodic function of 
$\theta$, i.e.
$E_{vac}(\theta\rightarrow 2\pi+\theta)=E_{vac}(\theta)$. This 
periodicity is a 
direct consequence of   quantization of the topological charge in QCD:
\bea
\label{2}
|\theta\ra=\sum_{n}e^{in\theta}|n\ra ,~~|\theta+2\pi\ra\equiv 
|\theta\ra ,
\eea
where $|n\ra$ is the winding state, $n$ is an integer. We  note
that the vacuum energy in gluodynamics (QCD without quarks) is  also
   a  $2\pi$ periodic function. \\
3. In gluodynamics with the large number of colors $N_c$
the physics should depend on $\theta$ through the combination
 $ \theta / N_c $ in order for the U(1) problem to be 
solved \cite{Wit}.\\
4. The anomalous WI's require \cite{Wit2} that Goldstone bosons 
which are described by the 
unitary matrix $U_{ij}$   should 
appear (apart from the mass term) 
 only in the following combination with $\theta$:
\bea
\label{3} 
  \theta - i \, Tr \, \log U .
\eea
This is the only allowed  combination  
for arbitrary $N_c$ and $N_f$. This fact is a consequence of 
the transformation
properties of the Goldstone fields $U\rightarrow 
\exp{i\alpha}U$ and the 
$\theta$ parameter   $\theta \rightarrow \theta +N_f\alpha$ 
under the $U(1)$ chiral rotations. 
 
\section{Di Vecchia-Veneziano-Witten (VVW) Solution }
Based on the results outlined in the previous section, and also
on the large $N_c$ arguments, 
Di Vecchia, Veneziano and Witten \cite{Wit2} suggested the 
following 
effective potential to analyze  the $\theta$ dependence 
of $E_{vac}(\theta)$:
\beq
\label{4}
W_{VVW}( \theta, U) = - \frac{ \la \nu^2 \ra_{YM} }{2} 
( \theta - i \log Det \, U )^2 
-\frac{1}{2}Tr \, (MU + M^{+}U^{+} ) + \ldots \; . ~~~~~~~
\eeq 
In this formula 
 $ M = diag (m_{i}  | \la \bar{\Psi}^{i} \Psi^{i} 
\ra | )$ and 
the coefficient in front of the combination
$ ( \theta - i \log Det \, U )^2$ is   
the topological susceptibility in pure YM theory rather than in 
real QCD:
$\la \nu^2 \ra_{YM}=-\frac{\partial^2E_{YM}}{\partial\theta^2} < 0$. 

The minimization of the potential $W_{VVW}( \theta, U)$ with 
respect
to the $U$ fields gives the  $\theta$ dependence of 
the vacuum energy. The obtained result has the required $2\pi$
periodicity and actually is a more complicated function
than the  simple ansatz $\cos(\theta)\sim \cos (a/ f_a)$
exploited in most calculations with the axions. 
%We should note that the shape and properties of 
%$E_{vac}(\theta)$ are very sensitive to the parameters
%of the potential $W_{VVW}( \theta, U)$, specifically
% to the vacuum condensate
% $\la \bar{\Psi}^{i} \Psi^{i} \ra  $
%which depends on temperature. For example, 
%according to (\ref{1}), the temperature dependence of the axon mass
%is the same as for the chiral condensate. 
 
In spite of the great simplicity and attractiveness of the 
VVW solution,
this scenario can not be quite complete. The argument is that 
pure YM theory, not only QCD,  should also obey   the
$2\pi$ periodicity law in $\theta$. For YM theory in the 
large $N_c $
limit the potential (\ref{4}) would imply
that the $\theta$ dependence in gluodynamics is given by
$W_{YM}( \theta)-W_{YM}( \theta=0) = - \la \nu^2 
\ra_{YM} 
\theta^2/ 2$, which can not be correct for an arbitrary $\theta$.
The first guess would be that our failure
to reproduce the $2\pi$ periodicity 
is related to shortcomings of the $1/N_c$ expansion:
 the next terms,
suppressed by factor $1/N_c$,  presumably should recover the 
periodicity. However, this  simple guess is not working. Indeed,
  any  function like $W_{YM}( \theta)=-
\frac{ N_c^2\la \nu^2 \ra_{YM}}{2}\cos(\theta/ N_c)$
which depends, as it should, on the 
combination $ \theta/ N_c$, would have period $2\pi N_c$ 
rather than the required $2\pi$. 
In addition, the approach of Ref.\cite{Wit2}  deals 
from the very beginning with the light chiral degrees of 
freedom and 
explicitly incorporates the $ U_{A}(1) $ anomaly without 
restriction
of the topological charge to integer values.  
 
Therefore, something is missing...
To understand exactly what is   missing, we should have a better 
understanding 
of   gluodynamics and its periodic properties.
 Only after that we can come back to QCD. As the first step in this 
direction
we would like to learn some lessons
about the $\theta$ dependence which supersymmetric (SUSY) models 
offer 
to us.

\section{Lessons from SUSY theories}

 In this section we would like to overview some SUSY models 
with an emphasize on   the properties of   
vacuum states and their $\theta$ dependence. As we already learned
in the case of QCD, this is the most relevant information for the 
axion physics.
Instrumental for this analysis is the 'old-fashioned' 
 effective Lagrangian approach, in which 
the effective Lagrangian is  defined   
as the Legendre transform of the generating functional
for connected Green functions.
Only the potential part
of this Lagrangian can be fixed in this way as it   corresponds to zero
momentum $n$-point correlation functions. The kinetic part is 
not fixed
in this framework. Thus, such an effective Lagrangian is useless
for calculating the $S$-matrix, but is perfectly suitable for   
  addressing the vacuum properties of the 
theory.   Fortunately, this is exactly the information we need:
the $\theta$ dependence of the vacuum energy is the
problem    amenable to a study within this framework.

To be more specific, let us consider the effective Lagrangian
for supersymmetric QCD (SQCD). The potential
part of the Lagrangian, 
 which is fixed by the anomalous Ward identities, is given 
by \cite{VY}:
\bea
\label{5}
\left. \left. W_{SQCD}= -\frac{1}{3}S \ln( 
\frac{S^{N_c-N_f}Det \, U}{\Lambda^{N_c}e^{-i\theta}}) \right|_F - 
Tr(m U) \right|_F , 
\eea
where $S$ is the gauge chiral superfield,
$\Lambda$ is the fundamental scale parameter of the theory,
 and $U^i_j
=Q^i\tilde{Q_j},~ i, j=1,..N_f$ 
is the matter superfield with the mass matrix $m$.
This Lagrangian should describe the vacuum structure of the
theory. As was found in \cite{SV}, there are  $N_c$
different vacua in SQCD, which 
are labeled by the $\theta$ angle and the 
 discrete parameter $k=0, 1,... N_c-1$
such that the gluino condensate depends
on these parameters as 
 $\la \lambda \lambda \ra\sim
\exp{(\frac{i\theta +2\pi k}{N_c})}$.
Therefore,  when $\theta $ varies 
continuously from $0$ to $2\pi$, $ N_c $ distinct and disconnected
Bloch type vacua undergo a cyclic permutation: 
 the 
first state
becomes the second one, and so on. All physical quantities are 
periodic in $ \theta $ with periodicity $ 2 \pi $, as these 
vacua can be just relabeled by the substitution   
$k\rightarrow k-1$ after the shift $ \theta \rightarrow 
\theta + 2 \pi$,   
keeping the physics intact \cite{SV}. 

Therefore, in SUSY theories we do have  the property
 described in Sect.2: the $\theta$ dependence 
comes through the combination $\theta/N_c$, but 
the physics is $2\pi$ periodic
due to the existence of the additional states. In
SUSY models they are degenerate, in non-supersymmetric theories
we expect that they have different   energies.

Although the potential (\ref{5}) has some appealing features, 
it can not be complete as was recently argued by Kovner and 
Shifman \cite{KS}. Indeed, the scalar potential
  (\ref{5}) is not a single-valued  function of the field.
If one starts at some $S=S_0$ and travels continuously 
to the different (but physically equivalent)   
$S_{0}'=e^{2i\pi}S_0$,
the value of the  potential at $S_{0}'$ will be different 
from that at
 $S_{0}$. Secondly, the discrete symmetry 
inherent to the original
theory is not reflected in Eq.(\ref{5}).
Both these shortcomings were successfully cured in \cite{KS}
 by the prescription
of summation over all branches of the multivalued potential (\ref{5}).
In the simplest case of supersymmetric
gluodynamics ($N_f=0$ in Eq.(\ref{5})), this prescription
leads to the following definition of the effective potential $W_{KS}$:
\bea
\label{6}
\exp (-W_{KS}) = \sum_n 
\exp \left[ \frac{1}{3}S \ln \left(
\frac{S^{N_c}}{\Lambda^{N_c}} \right) 
+h.c.+\frac{1}{3}(S-\bar{S})(\theta+2\pi n)
\right], 
\eea
where for simplicity we suppressed the integrals over the superspace
coordinates.
In terms of the original theory, 
such prescription means the summation over all
topological classes 
\bea
\label{7}
Z\sim\sum_n \int DA 
\exp \left[ -S_0+i\int d^4x(\theta+2\pi n)\frac{1}{32 \pi^2} 
G \tilde{G} \right]
\eea
which
 imposes  quantization of  the topological charge
$\int d^4x (1/32 \pi^2) 
G \tilde{G}$ in integer units. Therefore, the summation over all
topological classes does not change our theory, but simply introduces
an overall (infinite) factor into $Z$, which is  irrelevant anyhow.
  It is clear that all anomalous Ward identities,
as well as  the dynamical part of the effective lagrangian 
(the first term in Eq.(\ref{6}))  
 are kept intact by this prescription. 

It is not the purpose of this paper
  to discuss the technical details of this 
new development in gauge theories. We  refer the 
reader to the original papers
\cite{KS,HZ} for details.
Here we would like to end up this section by emphasizing  
some lessons
to be learned from SUSY theories, which are relevant to us and 
important for the axion physics:\\
1. $\theta\rightarrow\theta+2\pi$ is the explicit symmetry of 
the theory
because it  can be compensated for by a shift in $n\rightarrow n+1$
such that the potential (\ref{6}) is unchanged.\\
2. The vacuum states are classified by {\it two} parameters 
$|\theta, k=0,..N_c-1\ra$.
All these states are physically equivalent, but they are 
different vacuum states
labeled by the condensate  $\la \lambda \lambda \ra\sim
\exp{(\frac{i\theta +2\pi k}{N_c})}$.
This additional discrete quantum number $k$ can not be found 
from the analysis
of the symmetries of the original lagrangian because
of its quantum origin related to the  anomaly.
 It can be understood  only from explicit dynamical calculations.\\
3.  In SUSY models these extra states are degenerate in energy. 
However, 
in non-supersymmetric case one could generally expect that 
states with different energies (i.e. metastable
vacua) appear for some values of 
$\theta $, which decay by tunneling.
For the {\it axion physics}, this implies that 
decays of such metastable 
states may be supplemented by emission of the axions. \\ 
4. In general, there are domain walls in the system connecting 
different vacua. For the axion physics it would imply
the existence of different kind of  domain walls along with
the ones  studied previously \cite{dw}.

\section{Effective Lagrangian  and $\theta$ dependence in QCD}

In this section we describe the effective potential in QCD \cite{HZ}
which allows one to analyze the $\theta$ dependence of the ground state. 
In this approach, the Goldstone fields are 
described by the unitary matrix $ U_{ij} $ corresponding to the
$ \gmf $ phases of the chiral condensate: $ \la \bar{\Psi}_{L}^{i} 
\Psi_{R}^{j} \ra 
=  - | \la \bar{\Psi}_{L} \Psi_{R} \ra | \, U_{ij} $ with 
\beq
\label{u}
U = \exp \left[ i \sqrt{2} \, \frac{\pi^{a} \lambda^{a} }{f_{\pi}}  + 
i \frac{ 2}{ \sqrt{N_{f}} } \frac{ \eta'}{ f_{\eta'}}  \right] 
 , \;  
U U^{+} = 1 ,
\eeq
where $ \lambda^a  $ are the Gell-Mann matrices of $ SU(N_f) 
$, $ \pi^a $ 
is the pseudoscalar octet, and 
$ f_{\pi} = 133 \; MeV $.
In terms of $U$ and the ``glueball'' fields $h,\bar{h}$,
the QCD effective potential can be constructed analogously 
to the SUSY case, and has the explicit $2\pi$ periodicity: 
\beq
\label{8}
e^{-iV W(h, U) } =\sum_{n = - \infty}^{
 + \infty} \sum_{k=0}^{q-1} \exp \left[ -i V 
 W_d(h, U)  
+  i \pi V \left( k + \frac{q}{p} \,  
\frac{ \theta   + 2 
\pi n}{ 2 \pi}\right) \frac{h - \bar{h}}{2 i}\right].~~~   
\eeq
Here the ``dynamical" part $ W_d(h, U) $
 of the anomalous effective potential is   
\beq
\label{9}
W_{d} (h, U)  = \frac{1}{4} \frac{q}{p} h \, Log \left[ \left( 
\frac{h}{2 e E} \right)^{p/q}
  Det \, U   \right]    
- \frac{1}{2}Tr \, M U \; +h.c. 
\eeq
 where $V=\int d^4x$ is the 4-volume.
 All dimensional
parameters of this potential are expressed in terms of
the QCD vacuum condensates, and  are well known
numerically: $ M = diag (m_{i}  | \la \bar{\Psi}^{i} \Psi^{i} 
\ra | )$;  the constant $ E $ is related to the QCD gluon condensate 
$ E = \la b \alpha_s /(32 \pi) G^2 \ra $.
It is interesting to note that  the whole 
structure of Eq.(\ref{8}) is rather similar to that of the  
SUSY effective potential (\ref{6}). Namely, it contains
both the ``dynamical" $W_d$ and ``topological" parts (the 
first and the second terms in the exponent, respectively).
The ``dynamical" part of the effective potential (\ref{8}) 
is similar to   $W_{SQCD}$ 
  (\ref{5})
while the ``topological" part is 
akin to the improvement \cite{KS} of this effective potential. 

The only unknown parameters in this construction
are the integers $p,q$, which play the same role
as discrete integer numbers in SUSY theories, see Sect.4.  
These numbers are related to a discrete
symmetry which is a remnant of the anomaly,
and can be found only by explicit dynamical calculations.
    One can argue \cite{HZ1} that 
$q/p=8/3b$ where $b= (11/3) N_c - (2/3) N_f $
is the first coefficient of the $\beta$-function.
However, for more generality we prefer not  to fix $q/p$
in what follows with the only constraint that at large
$N_c,~~q/p\sim 1/N_c$ in order for the $U(1)$ problem to be solved. 
 
The heavy ``glueball" fields $ h , \bar{h} $ can be integrated 
out in Eq.(\ref{8}). The resulting
effective chiral potential is periodic 
in $ \theta $ and takes the form \cite{HZ} 
\bea
\label{11}
W_{QCD}(\theta,U,U^{+}) =  - \lim_{V \rightarrow 
\infty} \; \frac{1}{V} \log \left\{ 
 \sum_{l} \exp \left[ 
V E \cos \left[ - \frac{q}{p} ( \theta - i \log  Det \, U ) 
+ \frac{2 \pi}{p}
\, l \right]  \right. \right. \nonumber \\
\left. \left. + 
\frac{1}{2} V \, Tr ( M U + M^{+} U^{+} ) \right] \right\}   
\; \; , \; \; l= 0,1, \ldots , p-1 
%Z=e^{-VW_{QCD}(\theta, U,U^{+})} =  
% \sum_{l} e^{\left[-VW_{QCD}^l(U,U^{+})\right]} \nonumber\\
 % W_{QCD}^l=- E \cos \left[\frac{q}{p} ( \theta - i \log  Det \, U ) 
%-\frac{2 \pi}{p} l \right]      \nonumber \\
%  - \frac{1}{2}Tr ( M U + M^{+} U^{+} )~~     
% l= 0, \ldots , p-1 ~~.
\eea
 It was argued in \cite{HZ} that Eq.(\ref{11}) represents 
the 
anomalous effective Lagrangian realizing broken conformal and 
chiral symmetries of QCD. The arguments are the following: 
(a) Eq.( \ref{11}) correctly reproduces
 the VVW
 effective chiral lagrangian \cite{Wit2}
in the large $ N_c $ limit; (b) Eq.( \ref{11}) reproduces the anomalous 
conformal and chiral Ward identities of QCD.  
      
(a) For small
values of $ ( \theta - i \log  Det \, U ) < \pi/ q $,
the term with $ l = 0 $ dominates the infinite volume limit.  
Expanding the  cosine (this corresponds to the expansion
in $ q/p  \sim 1/N_c $), we recover
exactly  the VVW 
 effective potential (\ref{4})
  at lowest order 
in $ 1/N_c $, together with 
the constant  term $ - E = - 
 \la b \alpha_s /(32 \pi) G^2 \ra $ required by the conformal anomaly:
  \beq
\label{12}
W_{VVW}( \theta, U, U^{+}) = 
-E - \frac{ \la \nu^2 \ra_{YM} }{2} 
( \theta - i \log Det \, U )^2
- \frac{1}{2} Tr \, (MU + M^{+}U^{+} ) + \ldots \; . ~~~~~~~
\eeq
where  we used the fact that at large $N_c~ E(q/p)^2=
 -\la \nu^2 \ra_{YM} $ is the 
topological susceptibility in pure YM theory. 
Corrections in $ 1/N_c $ stemming from Eq.(\ref{11}) 
constitute a new result.  

(b) It is easy to check that the anomalous chiral and conformal 
WI's are reproduced by Eq.(\ref{11}).   As 
an important example, let us 
calculate the topological susceptibility 
in QCD near the chiral limit.
  For simplicity, we consider the limit of $ SU(N_f) $ 
isospin symmetry with $ N_f $ light quarks, $ m_{i} \ll 
\Lambda_{QCD} $.
For the vacuum energy for $ \theta < \pi/q  $ we obtain  
\beq
\label{13}
 E_{vac} (\theta) = -E + m_q N_f\lo\bar{\Psi}\Psi\ro
\cos \left( \frac{\theta}{N_{f}} \right)
\eeq
Differentiating this expression twice in $ \theta $, we reproduce
the famous WI (\ref{1}).

 We note that in general (Eq.(\ref{13}) is 
the particular example)
the $\theta$ dependence comes in a combination $\theta/N$
which naively does not provide the desired $2\pi$ periodicity
for the physical observables. At the same time, this 
formula was derived
from Eq.(\ref{11}) which is perfectly $2\pi$ periodic.
How   could it be?
 The answer is: the 
thermodynamic limit $ V \rightarrow \infty $ is performed for 
a certain value of $ \theta $, such that 
only a term of lowest energy survives in (\ref{11}),
 while all other states 
have higher energies and therefore drop out.
% in the standard
%definition
%\beq
%\label{14}
%E_{vac}( \theta) = - \lim_{V \rightarrow \infty} \frac{1}{V}
%\log Z \; \; , \; \;  \theta \; fixed \; 
%\eeq
%For the larger values of $\theta$ a different branch (different 
%value of the parameter $l$) gives the most important contribution
%for the lowest energy.
% On the other hand, due to the superselection rule
%different states do not communicate to 
%each other (and are absolutely
%stable), and therefore the fact of existence of additional
%higher energy states could be safely neglected, in
%agreement with Eq.(\ref{14}), for all physical
%problems except for one. Namely, retaining all these states 
%is necessary for the analysis of periodicity in $ \theta $. 
At the same time,
the values $ \theta $ and $ \theta + 2 \pi $ are physically
equivalent for the whole set of states and not for 
a selected individual vacuum state.
Thus, the fact that the $ \theta $ 
dependence in usual $ V = \infty $ 
formulas comes in the combination $ \theta/N $   
  has nothing to do with the problem of 
periodicity in $ \theta $, as those formulas refer to one 
particular state out of this set \cite{HZ}.

 \section{Applications to the axion physics}
\subsection{Local properties of the axion potential}

Now we are ready to apply our results
to the axion physics. The axion potential, by definition,
is obtained from Eq.(\ref{11}) by the replacement
$\theta\rightarrow a/f_a $:
\beq
\label{14}
W(a,U,U^{+})\equiv W_{QCD}(\theta=a/f_a, U,U^{+}).
\eeq
The next step is to integrate out the Goldstone fields $U$
exactly in the same way as was done in obtaining Eq.(\ref{13})
for small values of $\theta$. Technically, this problem amounts
to the minimization of the potential $W(a,U,U^{+})$ 
with respect to $U$ for a fixed value of $a$.

Before discussing the general global  properties of this potential, 
we would like
to emphasize that all parameters of the potential 
(except for the integers $p,q$) are fixed at zero temperature: 
$   \la \bar{\Psi}  \Psi \ra_{T=0} \simeq -(250 \, MeV)^3;~~
    \la  \alpha_s /\pi G^2 \ra_{T=0} \simeq 1.2 \cdot 10^{-2} \, 
GeV^4;~~
f_{\pi}(T=0)=133 \, MeV$. The dependence of these parameters
on temperature is also (at least, qualitatively) known.
In particular, the axion mass, which is defined 
as the quadratic coefficient in the expansion of the function
$E_{vac}(\theta)$ at small $\theta$, is proportional to
the chiral condensate: 
$m_a^2(T)\sim m_q \lo\bar{\Psi}\Psi\ro_T/ f_{a}^2$.
Therefore, $m_a^2(T)$ is known as long as $\lo\bar{\Psi}\Psi\ro_T$ 
is known. This statement is exact up to the higher order 
corrections in
$m_q$. We neglect these higher order corrections everywhere for
$T\leq T_c$ ($T_c\simeq 200 \, MeV$ is the critical temperature),
where the chiral condensate is nonzero and gives the most important
contribution to $m_a$. 
For the particular case 
 $N_f=2$
one expects the second order phase transition and, therefore,
$m_a^2\sim\frac{m_q}{f_a^2}\lo\bar{\Psi}\Psi\ro 
\sim |T_c-T|^{\beta}$
for   $T$  near $ T_c \simeq 200 \, MeV $. 
This is exactly where the 
axion mass 
does ``turn on''. The critical exponent in this case 
$\beta\simeq 0.38$,
see e.g. recent reviews \cite{smilga} for a general discussions 
of the 
QCD phase transitions. 

In what follows we make  the very plausible assumption
that the temperature dependence  for all   observables
for $T\leq T_c$  
comes entirely through the parameters  
 involved in the effective
 potential (the quark and gluon condensates and $f_{\pi}$). 
This assumption is based on the fact
 that the only relevant degrees 
of freedom at $T\leq T_c$ are the   
  Goldstone bosons. With this assumption, 
  the entire potential
$W(a,U,U^{+})_T$  for $T\leq T_c$ 
(not only the first term of its expansion $\sim m_a^2$) is 
also known.
Therefore, the procedure described above allows us to construct
the axion potential everywhere. In particular, for small
fields $a$ in the limit of $ SU(N_f) $ 
isospin symmetry with $ N_f $ light quarks, the potential is:
 \beq
\label{15}
 V(a) = -E + m_q N_f\lo\bar{\Psi}\Psi\ro_{T}
\cos \left( \frac{a}{f_aN_{f}} \right) +0(m_q^2),~~~
\frac{a}{f_a}\ll\pi,
~~~~~~~T\leq T_c .
\eeq 
Up to now we   discussed
only  the local properties of the axion potential. Its  
global (or topological) properties may be even 
more important for the axion physics.
 
\subsection{Global properties of the axion potential}

The topological 
properties of the potential
crucially depend on the integer numbers $q, p$ we introduced earlier.
Therefore, we consider the cases $q=1$ and $q\neq 1$ separately.
We start from the simplest $q=1$ case. In this case  
the global properties of the potential qualitatively are similar
to those of the VVW potential (\ref{4}). The only difference from
the latter is the presence of cusp singularities at certain values 
of the fields, which are a remnant of the topological charge 
quantization
in the effective Lagrangian approach \cite{HZ}.

An interesting qualitative 
phenomenon which follows from the analysis of the potential 
(\ref{4}) or 
(\ref{11}) 
  is a possible appearance  of   additional
 local minima 
at   $\theta \sim \pi$, depending on temperature and quark masses.
Therefore, the axion
potential in this region may become a
multi-valued function, i.e. there would be two different
values of the axion potential $V_{1,2}(\theta=a/f_a)$
for a fixed $\theta$, which differ by the phase of the chiral
field. 

In particular, for the VVW potential (\ref{4}) at $ T = 0 $ for 
three flavors with equal masses and $\theta=\pi$, there are two 
degenerate states \cite{Wit2} separated by the domain wall.
The wall surface tension in this case
was recently calculated by Smilga \cite{Smilga}:
 \beq
\label{16}
\sigma= 3\sqrt{2} \left(1-\frac{\pi}{3\sqrt{3}}
\right) m_{\pi}f_{\pi}^2 .
\eeq 
 For $\theta\neq\pi$ the energies of the vacua
are not degenerate anymore but are splitted apart by the amount
$\Delta E\sim m_q(\theta-\pi)$. For the case of equal masses, 
a metastable state in the VVW scenario exists in the region
 $ \pi/2\leq\theta \leq\ 3 \pi/2$ \cite{Smilga}. (As we will 
discuss shortly, this phenomenon becomes far more general
for the potential (\ref{11}) if $ q \neq 1 $.)
 
A metastable vacuum 
decays to the ground state with the formation of bubbles of the 
stable phase. The quasiclassical formula for the decay rate per 
unit time per unit 
volume was derived  many years ago\footnote{
This formula is derived for zero temperature, for a very 
high temperature \cite{Linde}
the correct expression is  
$ \Gamma\propto\exp\{- 16 \pi \sigma^3 /( 3 T(\Delta E)^2) \} $.}
 \cite{Okun}:
 \beq
\label{17}
\Gamma\propto\exp \left( -\frac{27\pi^2\sigma^4}{2(\Delta E)^3} 
\right) 
\equiv \exp ( - S_4 ) .
\eeq 
Using (\ref{16}), one finds that in the VVW scenario the lifetime
of the metastable state at zero temperature
is much larger than the
age of the Universe \cite{Smilga}. Such phenomenon
might play an important role in the development of the early
Universe during the QCD epoch.
 However, to make the corresponding estimates  
 one can not literally use
 Eq.(\ref{17}) because the parameters which
enter this formula  
depend on temperature and may drastically change the result
near the phase transition point.
It is quite possible that during the phase transition
the relevant factor which enters (\ref{17}) vanishes:
$ \sigma^4/(\Delta E)^3 \sim |T_c-T|^{\alpha}
\rightarrow 0,~~
\alpha >0$
at $|T_c-T|\rightarrow 0$ and, therefore,  $\Gamma\sim 1$. 
Indeed, in the mean field approximation where 
$\lo\bar{\Psi}\Psi\ro_T
\sim(T_c-T)^{1/2},~~
f_{\pi}^2\sim (T_c-T)$ we have $\sigma\sim (T_c-T)^{3/4}$ and 
$\Delta E\sim (T_c-T)^{1/2}$. Therefore 
$ \sigma^4 /(\Delta E)^3 \sim |T_c-T|^{3/2}\rightarrow 0$.
  
Now we assume that the axion exists. In this case,
if the axion field at temperature $T\simeq T_c$
is trapped in a metastable minimum at $\theta \sim \pi$,
    there are two   
different options for the decay of 
 this false vacuum. 
If $\Gamma\sim 1$ at $T\simeq T_c$,
tunneling is not suppressed, and the false vacuum tends to lower
its energy first by  
bubble nucleation without changing
$ \theta $, and then by relaxing 
to $ \theta = 0 $ with production of axions. This is advantageous
since the axion potential is nearly flat.   
However, if $\Gamma $ 
is still very small at $T\simeq T_c$,
the main mechanism for the decay could be related to the 
direct axion production.
 
For the case $q\neq 1$ (which we 
prefer, see \cite{HZ1}), the physics is even more interesting.
Additional metastable vacua appear 
for arbitrary quark masses and for an arbitrary $\theta$, 
not necessarily  in the region $\theta \sim \pi$. Therefore, 
decays of these false vacuum states will be a general phenomenon.
In this case the wall surface tension
can be easily calculated from (\ref{11}) and for $\theta=0$
is given by (see \cite{FHZ} for details)
\beq
\label{19}
\sigma=\frac{4 p}{q \sqrt{N_f} }  
f_{\pi}\sqrt{\la \frac{ b \alpha_s }{ 32 \pi} G^2 \ra }
\, \left( 1 - \cos \frac{\pi}{ 2 p} \right) +0(m_q f_{\pi}^2 ).
\eeq 
This formula substitutes Eq.(\ref{16})
describing the wall surface tension 
at $ \theta \simeq \pi $ for the VVW potential (\ref{4}).
A distinct difference between  these two cases is the absence
of the chiral suppression $\sim m_q$ in Eq.(\ref{19}),
which apparently 
would make penetration through the barrier even more difficult
in comparison to the VVW potential. The energy splitting
between the ground state and metastable state at $\theta=0$ is 
 \beq
\label{20}
 \Delta E =   m_q N_f \left| \la \bar{\Psi}\Psi\ra \right| 
\left( 1-\cos  \frac{2\pi}{qN_{f}} \right)  +0(m_q^2),
\eeq 
and we obtain
%\footnote{The $ N_c $ dependence displayed in 
%Eq.(\ref{21}) may look suspect as it apparently indicates
%that $ \Gamma \prop e^{-1/N_c} $ as $ N_{c} \rightarrow \infty $.
%Such a conclusion would be wrong,
%as the thin wall approximation implied in (\ref{17}) breaks down 
%at $ N_{c} \rightarrow \infty $ 
%for the potential (\ref{11}). Thus, in our case 
%the use of Eq.(\ref{17}) is unwarranted 
%if $ N_c $ is very large.}  
\beq
\label{21}
S_4 = 
% \frac{27\pi^2}{2} \frac{ \sigma^4}{2(\Delta E)^3} = 
\frac{ 3^3 \cdot 2^7 \cdot \pi^2 p^4}{q^4 N_{f}^5} \, \frac{f_{\pi}^4
E^2}{ M^3} \, \frac{ \left( 1 - \cos \frac{\pi}{2p} \right)^4 }{
\left( 1 - \cos \frac{ 2\pi}{q N_f} \right)^3 } \simeq 
\frac{27}{256} \, \frac{\pi^4 q^2 N_f}{p^4} \, \frac{
f_{\pi}^4  \la \frac{ b \alpha_s }{ 32 \pi} G^2 \ra^2}{ m_{q}^3 
\left| \la \bar{\Psi}\Psi\ra \right|^3 } \; . 
\eeq 
Eq.(\ref{21}) shows that the parametric suppression 
of the decay is largely overcome due to a numerical 
enhancement. The latter depends crucially on the particular 
values of the integers $ p , q $. In particular, for 
our favorite choice $ p = 11 N_c - 2 N_f , q = 8 $ \cite{HZ1},
Eq.(\ref{21}) yields a factor $ \simeq 10 $, while e.g. for 
$ p = N_c , q = 1 $ (as motivated 
by SUSY, see \cite{FHZ}) it is 
approximately two orders of magnitude larger,
but still much smaller than the estimate of \cite{Smilga} for 
the VVW potential. Of course, all remarks made above concerning 
the temperature dependence 
of such effect apply for the estimate (\ref{21}) as well.
In particular, 
 $\Gamma$ could be of order
one during the QCD phase transition 
in development  of the early Universe and, therefore, 
a metastable state could decay just before 
the Universe cools down to the temperature
where Eq.(\ref{21}) could be applied. 

Another related question is the dynamics of the 
domain walls which separate two different vacua
with the small splitting $\Delta E $ (\ref{20}).
These domain walls are absolutely harmless
for development of our universe because they
decay in proper time scale. Indeed, 
the required value for the pressure (which is equal
to $\Delta E $ in our notations)
for the safe decay of the wall is given by \cite{Vilenkin} 
\beq
\label{22}
 \Delta E =   m_q N_f \left| \la \bar{\Psi}\Psi\ra \right|
\left( 1-\cos  \frac{2\pi}{qN_{f}}  \right) 
\geq \frac{\sigma^2}{M_P^2},    
\eeq 
where $M_P$ is the Plank scale.
Inequality (\ref{22}) is perfectly satisfied with our parameters
$\Delta E $ and $\sigma$ (\ref{19}). 
Therefore, such domain walls do not lead to a
cosmological disaster, but rather 
 may have very interesting cosmological consequences 
which still have to be explored.
Similar domain walls which separate
vacua with different phases of the gluino condensate
have been  recently discussed by Shifman
in SUSY models \cite{Shif}. It is quite remarkable 
that this phenomenon may exist not only in SUSY models but also
in the physically relevant case of QCD.

We conclude this section by emphasizing that the phenomena
just described may lead to the 
necessity  to reconsider the  constraints on a dark
matter axion because the dynamics of the axion field
could be more complicated than it was originally thought.
We stress  
that the  decay of the metastable state
described above  proceeds (with or without axions) through the
expansion of bubbles with $100\%$ violation of CP invariance. 
This is because the phase of the chiral condensate
in the metastable vacuum is nonzero and of order 1. This 
leads to violation of CP even if $ \theta = 0 $.
(This is not at variance with the Vafa-Witten theorem \cite{VW}
which refers to the lowest energy state only.) 
It may have 
profound consequences for the development of the early Universe
at the QCD scale because such effects could lead to 
a new mechanism
for baryogenesis! Indeed, the famous Sakharov criteria \cite{Sakh}
could be satisfied in this scenario, see Sect.8 
for a more detailed  discussion.

\section{ New axion search experiment at RHIC?}

The development of the early Universe is a 
 remarkable laboratory for the study of most nontrivial
properties of the particle physics.
What is more amazing is the fact that these 
phenomena at the QCD scale can be,
in principle,  experimentally tested at RHIC, Brookhaven.

We expect that, in general, an arbitrary $|\theta\ra$-state
would be created in the heavy ion collisions,
similarly to the creation of the disoriented chiral condensate (DCC)
with an arbitrary isospin direction. It should be a large 
domain with a wrong
$\theta\neq 0$ orientation. As we shall see in a moment,
 for both cases (DCC and $|\theta\ra$-state) the difference in energy 
between a created state and the lowest energy state is proportional
to $m_q$ and negligible at high temperature. Therefore, 
energetically an arbitrary $|\theta\ra$ can be formed.
The way of how the created  $|\theta\ra$-state will  relax
to the ground state of lowest energy $|\theta =0\ra$
 is a separate issue. 
 If, somehow,
 an equilibrium state with a large correlation length 
in the large volume $V$ is formed,
the only possible way to relax $\theta$ to zero would be the 
axion mechanism (if they exist). 
However, due to the fact that we do not expect to create 
an equilibrium state with an infinite  correlation length 
in the heavy ion collisions,
 the decay of a $|\theta\ra$-state will also  occur due to the  
 Goldstone $U$ fields with specific $CP$-odd correlations\footnote{
A similar phenomenon has been recently discussed in Ref.\cite{Pisarski}}.
Therefore, two mechanisms of the relaxation of a $|\theta\ra$-state
to the vacuum would compete: the axion one and the standard decay
to the Goldstone bosons. In the large volume limit  if 
 a reasonably good equilibrium state 
with a large correlation length is created,
 the axion mechanism
would win; otherwise, the Goldstone mechanism would win. In any case, 
the result of the decay of a $|\theta\ra$-state
would be very different depending on the presence or 
absence of the axion
 field  in Nature.

Before   going  into details, we would like to recall
some general properties of the DCC which (hopefully) can be 
produced at RHIC
(see e.g. \cite{DCC} for a review),
with an
emphasize on the analogy between the DCC and a misaligned 
$|\theta\ra$-state.
If the cooling process is very rapid and, therefore, the 
system is out
of equilibrium, there will be  a large size of the correlated 
region
 in which the vacuum condensate orientation mismatches 
its zero temperature value.
The absolute value of the chiral condensate right after the 
phase transition is expected to  be close to its final
(zero temperature) magnitude. However the vacuum direction of 
the formed condensate
is still misaligned since it takes a longer time for the vacuum 
orientation
to relax due to the small free energy difference $\sim m_q$
between the formed and  true vacuum states.

To be more specific, let us consider the case $ N_f = 2 $.
The matrix $ U $ is parametrized by the misalignment angle $ \phi $
and the unit vector $\vec{n}$ in the isospin space: 
 
\beq
\label{23}
U=e^{i\phi(\vec{n}\vec{\tau})} \; , \; Tr(\tau^a\tau^b)=2\delta^{ab} \; 
, \; 
U U^{+} = 1 \; , \; \la \bar{\Psi}_{L}^{i} 
\Psi_{R}^{j} \ra 
=  - | \la \bar{\Psi}_{L} \Psi_{R} \ra | \, U_{ij}
\eeq
The energy density of the DCC is determined by the mass term:
\beq
\label{24}
 E_{\phi}= -\frac{1}{2} Tr( M U + M ^{+}U^{+}) = 
 - 2m  | \la \bar{\Psi}  \Psi  \ra | \cos(\phi)
\eeq
where  we put $m_u=m_d=m$  for simplicity.
Eq.(\ref{24}) implies that any $\phi\neq 0 (mod ~2\pi)$ 
is not a stable vacuum state because $\frac{\partial
 E_{\phi}}{\partial\phi}|_{\phi\neq 0}\neq 0$, i.e. the vacuum 
is misaligned.
On the other hand, the energy 
difference
between the misaligned state and true vacuum 
with $\phi=0$ is small
and proportional to $m_q$. Therefore, the probability to create 
a state
 with an arbitrary $\phi$ at high temperature 
$ T \sim T_c $ is proportional to
$ \exp[- V( E_{\phi} - E_0)/T]$  and  depends on 
$\phi$ only very weakly, i.e. $ \phi $ is a quasi-flat 
direction. Right after the phase transition when $\la 
\bar{\Psi}  \Psi  \ra $ 
becomes nonzero,
the pion field begins to roll toward $\phi=0$, and of course 
overshoots $\phi=0$.
Thereafter, $\phi$ oscillates. One should expect the coherent 
oscillations of the 
$\pi$ meson field which would correspond to a zero-momentum 
condensate of pions.
Eventually these classical oscillations produce real $\pi$ mesons which
hopefully can be observed. In a sense this picture is very similar to
the standard ``misalignment'' mechanism \cite{misal,review}
for production of the cosmic axions during the QCD epoch.

Now we turn to our main point when the $U(1)_A$ phase of the disoriented 
chiral 
condensate is also nonzero  and, therefore, the $|\theta\ra$-vacuum 
state
could be formed. To take into account this $U(1)_A$ phase we choose 
the matrix $ U_{ij} $ in the form 
$U=diag \; (e^{i\phi_i} ) $.
The energy density of the misaligned vacuum is determined in this 
case by
Eq.(\ref{11}). The most important difference 
between
Eqs. (\ref{24}) and (\ref{11}) 
is 
the presence of the parametrically large term  $ \sim E\gg m_q 
| \la \bar{\Psi}  \Psi  \ra |
$ in the 
expression for energy (\ref{11}), 
describing the $U(1)_A$ phase of the disoriented chiral condensate. 
This term, as was explained in Sect.5, is related to the 
anomalous WI's, and does not go away in the chiral limit. 

The key point is the following. For
arbitrary phases $ \phi_i$  the
energy of a misaligned state differs by a huge 
amount  $\sim E $ from the vacuum energy. 
Therefore, apparently there are no quasi-flat directions 
along $\phi_i$ coordinates, 
which would lead to the long wavelength oscillations with production
of a large size domain. However, when
the relevant combination  $(\sum_i\phi_i-\theta)$ from Eq.(\ref{11}) 
is close by an amount $ \sim O(m_q) $ to its vacuum value, a 
Boltzmann suppression due to the term $ \sim E $ is 
absent, and an arbitrary misaligned $|\theta\ra$-state
can be formed. In this case for any $ \theta $ 
the difference in energy
between the true $|\theta\ra$ vacuum
 and a misaligned $|\theta\ra$- state  (when the $\phi_i $
fields are not yet in their final positions 
$\overline{\phi_i(\theta)}$)
 is proportional to 
$m_q$ and very small in close analogy to the DCC case.

After this point we can apply
the same philosophy as for DCC.  
 The chiral fields\footnote{If 
$\theta\neq 0$,
the Goldstone fields are not exactly the pseudoscalar fields, 
but rather
are mixed with the scalars; the mixing angle between the singlet and 
octet  combinations
also depends on $\theta$,
see \cite{FHZ} for detail.}
$ \phi_i $
begin to roll toward the true solution $\overline{\phi_i(\theta)}$  
  and of course overshoots  it. The situation is very similar to what 
was described for the DCC with the only difference that  in general 
we expect an 
 arbitrary $|\theta\ra$-disoriented state to be created 
in heavy ion collisions,
not necessarily the $|\theta=0\ra$ state. The difference in energy 
between these states
is proportional to $m_q$, as follows from the fact that 
the $\theta$ 
dependence
of any physical observable is proportional to $m_q$. 

If  a reasonably stable  $|\theta\neq 0\ra$ state is created, 
it could decay into the axions\footnote{The possibility
of a production of the axions in the heavy ion collisions
was independently discussed by Melissinos \cite{ion}. 
AZ thanks Adrian Melissinos for a conversation on 
the subject.}.
 To estimate the effect, we consider an ideal case when
all energy  stored in the $|\theta \neq 0\ra$ state will be 
released through
production of the axions\footnote{We are, of course, aware
that this ideal case can not be realized in Nature.
There should be a strong suppression factor $\rho$ in front 
of Eq.(\ref{25})
due to the fact that there is no superselection rule in
a finite volume nonequilibrium state.
Therefore, the $\theta$ parameter can be relaxed to zero by others 
means, not necessarily
related to the axion productions.  The parameter
$\rho$ should depend, first of all, on the correlation 
length $\lambda$
of the formed misaligned $| \theta\ra$ state.
We do not know at the moment  
how to estimate this parameter $\rho(\lambda)$. For the
ideal case, $\rho=1$.}. In this case  from Eq.(\ref{13}) 
we expect the following axion density $n_a$ for a random parameter 
$\theta\sim 1$:
\beq
\label{25}
 n_a = \rho\, \frac{m_q N_f}{m_a} |\la \bar{\Psi}\Psi\ra|
\left(1-\cos   \frac{\theta}{N_{f}}\right)\sim \rho \cdot 10^{19}\, 
MeV^3
\left(\frac{10^{-5}\, eV}{m_a}\right),
\eeq
where parameter $\rho$ is a suppression factor, see footnote (5).  
The standard way to detect the produced axions  is to use their
property of conversion to photons
in an external  magnetic field \cite{sikivie}. The relevant 
axion-photon coupling constant is defined in the following way:
\beq
\label{26}
L_{a\gamma\gamma}=\frac{g_{a\gamma\gamma}}{4}a\tilde{F_{\mu\nu}}
F_{\mu\nu},~~
g_{a\gamma\gamma}(DFSZ)=\frac{\alpha}{2\pi f_a} 
~~~g_{a\gamma\gamma}(KSVZ)=-\frac{\alpha}{2\pi f_a}\cdot\frac{5}{3}
\eeq
where we specified the coupling constants in the limit $m_u=m_d$ for 
two 
popular models \cite{DFSZ}. 
The conversion probability  is given by \cite{sikivie}:
\beq
\label{27}
P_{a\rightarrow\gamma}=\frac{1}{4}g_{a\gamma\gamma}^2 \left( 
B^2l^2\right)
\eeq
where $B$  is the external magnetic field and  $l$ its length.
With our estimate (\ref{25}) for the axion density,
one can get the following formula for the 
probability of  detection of a photon from 
the axion conversion \cite{ion}: 
\beq
\label{28}
P_{\gamma}\sim
P_{a\rightarrow\gamma}n_aV\sim
\rho m_a (\frac{\alpha}{2\pi})^2 \left( B^2l^2\right) V,
\eeq
where we took into account the relation 
$f_a^2m_a^2\sim m_q\lo\bar{\Psi}\Psi\ro$,
see Sect.6.1.
In this formula $V \sim \lambda^3$ where $ \lambda$  is the  
correlation length 
 for the misaligned $|  \theta\ra$-state.
As was discussed above, the mechanism of production of axions
by the coherent processes corresponds to a condensate of 
the ``almost'' zero-momentum
($\sim m_q/\lambda$) axions.   If the collision is asymmetric (as 
was suggested in \cite{ion}), the photon signal from
the axion-photon conversion  will be unambiguous because it will be 
the ``almost'' monochromatic wave 
with $\Delta\omega\sim m_q/\lambda$ and 
$\omega$ to be determined by the  asymmetry in energies  of  the  
beams.
 
Therefore, the heavy ion collisions give us a unique
chance for a new axion search experiment.
We would like to stop here  with this optimistic note. 

\section{Conclusion}

$~~~\bullet$ Probably the most important outcome 
of our study is a better
understanding of the global and local properties of the 
axion potential $V(a)$, which may result in the necessity to reconsider 
the constraints on a dark
matter axion. The temperature dependence of the potential 
at $T\leq T_c$ 
is expressed  in terms of 
the $T$-dependent vacuum condensates which have been extensively
studied on the lattice and in the models \cite{smilga}
 and are fairly known.  Therefore, we know the temperature 
dependence of the
potential as long as we know the temperature dependence of 
the chiral and gluon condensates. 

$\bullet$ The axion potential
  is generically a  multi-valued function.  Such a property 
may lead to the phenomenon of the false vacuum 
decay through bubble nucleation,  
which may make the dynamics
of the axion field far more complicated (and interesting)
than was previously thought.
 
$\bullet$ Based on the analysis of the potential $V(a)$
expressed in terms of the condensates and the knowledge
of their temperature dependence,  we suggested 
a new idea for the axion search experiment at RHIC. 

 $\bullet$ As a  byproduct
  of our analysis (which was mentioned at the end 
of Sect.6), we would like to speculate on 
the possibility of baryogenesis at the QCD scale
(with or without axions). Indeed, it appears that
all three famous 
Sakharov criteria \cite{Sakh} could be 
satisfied in the decay of a metastable
state discussed above:\\
1. Such a metastable state is clearly out of 
equilibrium; \\
2. CP violation is unsuppressed and proportional
to $  m_um_dm_s \bar{\theta}_{eff}
 ,~\bar{\theta}_{eff} \sim 1$. 
As is known, this is the
most difficult part to satisfy
in the scenario of baryogenesis at the electroweak scale
within the standard model for CP violation; \\
3. The third Sakharov criterion 
is violation of the baryon $(B)$ number.
Of course, the corresponding U(1) is an exact global symmetry
of QCD. However, a ``spontaneous" breaking of the baryon
U(1) symmetry (as  
a result of interactions with the domain wall) is
not forbidden, and would be sufficient. The latter may 
arise in a way similar to electroweak baryogenesis,
see e.g. \cite{dolgov} for a review. 

 The viability and details of such a mechanism for baryogenesis
are still to be explored \cite{BHZ}, however,
 in general, one could expect a large
asymmetry according to the unsuppressed 
CP violation
$\sim  m_u m_d m_s / \Lambda^3$ in bubble nucleation. 
%which 
%is much larger 
%  the observed ratio $\frac{n_B-n_{\bar B}}{n_{\gamma}}\simeq 
%3\cdot 10^{-10}.$
%While in most scenarios the asymmetry is smaller than the 
%observed value and
%some attempts should be taken to make it as large as 
%possible, here the situation is reverse: 
%the asymmetry is expected to be very large 
%according to the measure of CP violation, and one 
%should invent 
%a mechanism to suppress it by a few order of magnitudes. 
What is  amazing is the fact that each step in
 such a scenario for baryogenesis could be,
in principle, experimentally tested at RHIC, Brookhaven.

 $\bullet$ The remark that QCD scale could
be an interesting    place to look
  is also motivated by  
recent observations of the MACHO collaboration \cite{macho}.
In particular, the  domain walls  with QCD scale
which always separate
the true and false vacuum states
do not lead to a
cosmological disaster, see (\ref{22}).  They rather 
 may be very interesting objects for the  problem
of the structure formation at small scales $\sim M_{\odot}$.

 \end{document}